\documentclass[fleqn,10pt]{wlscirep}
\usepackage[utf8]{inputenc}
\usepackage[T1]{fontenc}
\usepackage{listings}
\usepackage{algorithm}
\usepackage{amsmath}
\usepackage[english]{babel}
\usepackage[noend]{algpseudocode}

\usepackage{lineno}
\linenumbers

\title{Resonant Scanning Design and Control for Fast Spatial Sampling}

\author[1,*]{Zhanghao Sun}
\author[1]{Ronald Quan}
\author[1]{Olav Solgaard}

\affil[1]{Stanford University, Electrical Engineering, Stanford, CA, 94305, US}

\affil[*]{zhsun@stanford.edu}

\begin{abstract}
Two-dimensional, resonant scanners have been utilized in a large variety of imaging modules due to their compact form, low power consumption, large angular range, and high speed. However, resonant scanners have problems with non-optimal and inflexible scanning patterns and inherent phase uncertainty, which limit practical applications. Here we propose methods for optimized design and control of the scanning trajectory of two-dimensional resonant scanners under various physical constraints, including high frame-rate and limited actuation amplitude. First, we propose an analytical design rule for uniform spatial sampling. We demonstrate theoretically and experimentally that by expanding the design space, the proposed designs outperform previous designs in terms of scanning range and fill factor. Second, we show that we can create flexible scanning patterns that allow focusing on user-defined Regions-of-Interest (RoI) by modulation of the scanning parameters. The scanning parameters are found by an optimization algorithm. In simulations, we demonstrate the benefits of these designs with standard metrics and higher-level computer vision tasks (LiDAR odometry and 3D object detection). Finally, we experimentally implement and verify both unmodulated and modulated scanning modes using a two-dimensional, resonant MEMS scanner. Central to the implementations is high bandwidth monitoring of the phase of the angular scans in both dimensions. This task is carried out with a position-sensitive photodetector combined with high-bandwidth electronics, enabling fast spatial sampling at $\sim 100$Hz frame-rate.
\end{abstract}

\begin{document}

\flushbottom
\maketitle
\thispagestyle{empty}

% \vspace{-0.5cm}% Additional space between abstract & rest of document
\section*{Introduction}
% \label{sec:intro}
Recent years have seen the rapid development of LiDAR systems in robotics~\cite{robot}, autonomous vehicles~\cite{KITTI, lidar1, mems_lidar} and AR/VR applications~\cite{XR}. Designing such systems require innovation in both hardware and software because real-time response requires fast information collection and processing. Optical scanners are commonly used in LiDAR to deflect laser beam(s) onto different sampling positions in space and acquire 3D data. Compared to conventional LiDAR scanners that operate in a raster scanning mode, redresonant scanners have a well-known advantage~\cite{mems_lidar, nathan, microscopy1, zhang2010}: the motion amplitude of a resonant scanner is $\sim Q$ times larger than that of a raster scanner, where $Q$ is the quality factor of the resonant system~\cite{nathan, mems_lidar}. Resonant scanning also improve acquisition speed~\cite{high_fps1, high_fps2, high_fps3}. Raster scanners acquire data in a prescribed sequential pattern that is limited by the speed of its slow axis. This results in slow spatial sampling that is unacceptable for many real-world applications, e.g. collision avoidance. In contrast, resonant scanners have high speed in both scanning axes, which is promising for high-speed information collection. To realize this advantage, resonant scanning patterns must be optimized such that information is acquired most efficiently within a short frame time, and their flexibility should be increased to allow situation-dependent, or "random" scanning patterns~\cite{mirrorcle1, glv1, glv2}.

Multiple scanning pattern designs have been proposed~\cite{hdhf1, hdhf2, high_fps1, high_fps2, optim1, liss_design1}. Hwang et al. proposed a frequency selection rule for high frame-rate $\sim 10-100$Hz operation~\cite{hdhf1, hdhf2}. Tuma et al. applied optimization-based scanning trajectory design in scanning probe microscopy~\cite{optim1}, which operates at a lower frame-rate $\sim 1$Hz. Sub-frame sampling and updating were also proposed to boost the imaging updating rate~\cite{high_fps1, high_fps2, nathan, adaptive_liss}. These designs tend to focus on actuation frequency selection while ignoring the phase. More recent work discussed both frequency and phase in scanning pattern design~\cite{liss_design1}, but they are limited to patterns that repeat in each frame. Moreover, none of these designs considered physical constraints such as actuation signal amplitude, which are important in real-world systems. The flexibility in scanning is also critical. As shown in reports on random-access scanning~\cite{adaptive_sample1, adaptive_liss, glv1}, scanning patterns that ``focus'' on specified Regions-of-Interest (RoI) meet data post-processing requirements better than uniform spatial sampling. Resonant scanners cannot abruptly change direction so truly random-access scanning is not possible, and traditionally resonant scanning has been optimized for uniform Field-of-View (FoV) coverage.  Therefore, there is a need for approaches that allows RoI focused sampling using resonant scanners. 

In this work, we demonstrate optimized designs for resonant scanning patterns with frame-rate $\sim 100$Hz and limited actuation amplitudes. We first analyze uniform spatial sampling and introduce two metrics: fill-factor and scanning range. We show a trade-off between these two metrics in previous designs to motivate a better solution. An analytical design rule based on unmodulated scanning patterns (both axes have single-tone scanner motion) is proposed that takes various practical considerations into account, such as high frame-rate, bounded actuation amplitude, scanner phase, and pattern repeating period. The proposed design out-performs previous designs that fail to consider these factors. Furthermore, we consider RoI-focused spatial sampling with resonant scanners. For this purpose, we demonstrate the utility of modulated resonant scanning patterns, which contain multiple frequency components around resonance. We develop a task-driven optimization framework to integrate scanning pattern design with post-processing on sampled 3D data. 

To demonstrate the applications of designed resonant scanning patterns, we evaluate them in simulated 3D computer vision tasks including LiDAR odometry and object detection~\cite{lidar1, odom2, pointnet, fpointnet} (section ``Simulations''). To experimentally implement the designed patterns, we built a hardware prototype based on a MEMS scanner (section ``Experiments''). We developed a control system that stabilizes the scanner phase during operation, which is critical for resonant mechanical systems~\cite{robot_control, pll1, pll2, sandy_thesis, dual-tone, zhang2010}. Compared to previously designed high-accuracy, narrow bandwidth phase control systems, the proposed method is wide-band and thus can operates in both unmodulated and modulated scanning modes.

\section*{Scanning Pattern Design}
%  \label{sec:design}
Laser beams reflected from scanners that are resonant in two orthogonal dimensions create "Lissajous patterns" that are described mathematically as follows: 
\begin{equation}
\label{eqn:motion1}
    \left\{ \begin{array}{l}
    x(t) = A_x(t)cos(2\pi f_xt + \phi_x(t))\\
    y(t) = A_y(t)cos(2\pi f_yt + \phi_y(t))\\
    \end{array} \right.
\end{equation}
where $f_x$, $f_y$ are the scanning frequencies, which are assumed to be close to resonant frequencies ($f^r_x$, $f^r_y$). The quantities $A_x(t)$, $A_y(t)$ and $\phi_x(t)$, $\phi_y(t)$ are the amplitudes and phases for the two scanning axes. When amplitudes and phases are static, both $x(t)$ and $y(t)$ are single-tone, and we denote the corresponding scanning patterns as unmodulated patterns. When small modulations (or, equivalently, multiple frequency components within resonance bandwidth) are added, we denote the corresponding scanning patterns as modulated patterns. To make the problem of optimizing the scanning patterns tractable, we make the following assumptions:
(1) The amplitude of the actuation signal is bounded to reflect limitations on practical hardware. (2) We define a ``frame time'' $T_{frame}$ (Note that this ``frame time'' is different from that used in previous literature~\cite{high_fps1, high_fps2}, we provide a comparison between these two concepts in Supplementary Information). Data collected within $T_{frame}$ is used for evaluation or post-processing. We show that the bounded actuation amplitude and short frame time introduce a trade-off between two important metrics for spatial coverage: scanning range and fill factor, and motivates a better design rule. For simplicity, we choose $T_{frame} = m$, $m$ being an integer between $6$ to $9$ (corresponding to $6-9$ms $T_{frame}$ with a 1kHz resonant scanner). (3) Without loss of generality, we assume resonant frequencies for the two scanning axes to be $f_x^r = r$, $f_y^r = 1$. In all the simulations, time is scaled by the scanning cycle in the y-axis and is dimensionless. To model typical MEMS scanners, we set a quality factor $Q = 20$. We also limit $r \in [1,3]$. When $r$ gets higher, the resonant scanning system gradually transitions to a raster-scanning system, with one axis scanning much slower than the other.

\subsection*{Unmodulated pattern design}
% \label{sec:basic_design}
We first analyze the spatial coverage of resonant scanning with bounded actuation amplitude. The goal is to achieve uniform spatial sampling in a normalized $[-1,1]\times[-1,1]$ Field-of-View (FoV). We use fill-factor and scanning range to characterize the spatial coverage of scanning patterns, following common usage in the literature~\cite{hdhf1, liss_design1}. The fill factor characterizes the spatial coverage within a normalized scanning range. The range is determined by the scanning frequencies $f_x$, $f_y$ and the transfer function of the resonant scanner $H_x(f_x)$, $H_y(f_y)$. We quantify the two metrics in Equation~\ref{eqn:metric}:
\begin{equation}
\label{eqn:metric}
\left\{ \begin{array}{l}
\textrm{fill-factor} \triangleq 2 - R_{max}, \;\;\;\; \textrm{higher is better}\\
\textrm{scanning range} \triangleq H_x(f_x) \times H_y(f_y), \;\;\;\; \textrm{higher is better}\\
H_x(f_x) = 1 /(\sqrt{((f_x/f_x^r)^2 - 1)^2 + (f_x/(f_x^r Q_x))^2)} Q_x)\\
H_y(f_y) = 1 /(\sqrt{((f_y/f_y^r)^2 - 1)^2 + (f_y/(f_y^r Q_y))^2)} Q_y)
\end{array} \right.
\end{equation}

\begin{figure}[t!]
\begin{center}
% \fbox{\rule{0pt}{3in} \rule{0.9\linewidth}{0pt}}
\includegraphics[width=0.95\linewidth]{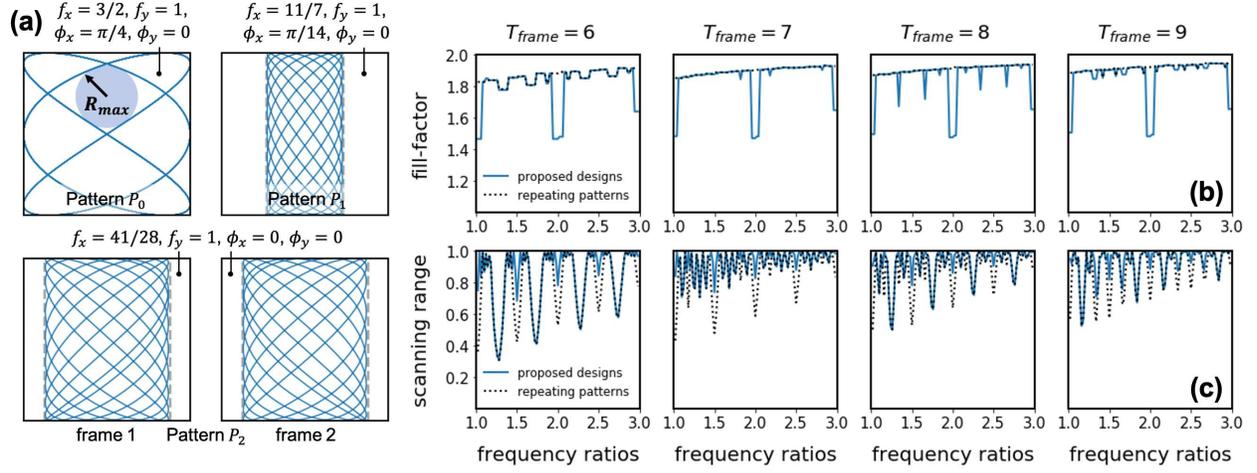}
\end{center}
   \caption{(a) Several scanning patterns for a resonant scanner with $f_x^r = 1.5$, $f_y^r = 1.0$ and $T_{frame} = 7$. Pattern $P_0$ is the on-resonance actuated pattern, Although it has large scannning range $1.0$, fill-factor is low ($0.63$), as indicated by the radius of its largest inscribed circle. Pattern $P_1$ uses $f_x = 11/7$, $f_y = 1$, $\phi_x = \pi/14$, $\phi_y = 0$ as parameters. The fill-factor is improved to $0.89$ but scanning range is reduced to $0.45$. Pattern $P_2$ uses $f_x = 41/28$, $f_y = 1$, $\phi_x = 0$, $\phi_y = 0$, derived from design rule~\ref{alg:design_single}. It has fill-factor $=0.88$ and scanning range $= 0.74$. (b),(c) Fill-factor/Scanning range with proposed design rule~\ref{alg:design_single} (blue, solid) and repeating-pattern design rule~\cite{liss_design1} (black, dashed), with different resonant frequency ratio and $T_{frame}$ settings.}
\label{teaser}
\end{figure}

We normalize the transfer function amplitudes to $1$. Note that we only consider the amplitudes of the transfer functions because in the proposed control scheme, we directly monitor phase of the scanner motion, instead of the phase of the actuation signals. Without loss of generality, we assume an ideal harmonic oscillator model for the resonant-scanner because the following analysis is only based on the band-pass characteristic, which are common to all resonant scanners. We also ignore cross-talk between x and y-axis motions in this simplified model. More discussions on cross-talk are presented in the discussion section. Similar to previous literature~\cite{optim1, liss_design1}, fill-factor is defined through the radius of the largest inscribed circle $R_{max}$ in the sampling pattern, as shown in Figure~\ref{teaser}(a). To decouple the two metrics, the scanning pattern is normalized to $[-1,1]\times[-1,1]$ when calculating $R_{max}$ (See Supplementary Information for details).

With bounded actuation amplitude, there is a fundamental trade-off between fill-factor and scanning range. As an example, we show several scanning patterns with $f_x^r = 1.5$, $f_y^r = 1$, $T_{frame} = 7$ in Figure~\ref{teaser}(a). If we actuate on-resonance, the scanning range is at maximum ($=1.0$), as shown in pattern $P_0$. $P_0$ repeats in $t = 2$ with $\phi_x = \pi/4$, $\phi_y = 0$ and it samples on exactly the same trajectory multiple times within $T_{frame}$. This results in a low fill-factor $= 1.63$. On the other hand, if actuated off-resonance, with $f_x = 11/7$, $f_y = 1$, $\phi_x = \pi/14$, $\phi_y = 0$, the sampling pattern has high fill-factor $= 1.89$ (pattern $P_1$). However, off resonance actuation leads to large reduction in scanning range ($= 0.45$). In the lower part of Figure~\ref{teaser}(a), we show another design (pattern $P_2$) with $f_x = 41/28$, $f_y = 1$, $\phi_x = \phi_y = 0$. $P_2$ has repeating period $14 = 2T_{frame}$. In each frame, $P_2$ has fill-factor $1.88$ and scanning range $0.74$. The fill-factor is almost the same as that in $P_1$ and the scanning range ($= 0.74$) is $1.6\times$ larger. Therefore, with respect to spatial coverage within $T_{frame}$, we regard $P_2$ to be a better scanning pattern when compared to $P_1$. 

Previous resonant scanning pattern designs generally consider patterns that repeat in each $T_{frame} = m$, with $f_x = k/m$, $f_y = l/m$, $k, l \in \mathbb{Z}$~\cite{hdhf1, liss_design1} and we denote these ``repeating patterns''. (Note that the ``repeating''/``non-repeating'' pattern definition here is different from the definition used in some previous literature~\cite{high_fps1, high_fps2}, we provide a comparison between these two concepts in Supplementary Information) However, for a specific $f_x^r$, $f_y^r$, $T_{frame}$ combination, there might not be a repeating pattern with ($f_x$, $f_y$) close enough to resonance. As in the example in Figure~\ref{teaser}(a), it can be verified that $P_1$ is the repeating pattern with ($f_x$, $f_y$) closest to resonance, but $P_1$ still suffers from small scanning range. In this paper, we expand the design space by considering not only repeating patterns, but also patterns with repeating periods longer than $T_{frame}$, such as $P_2$ in the above example. We propose an analytical design rule in design rule~\ref{alg:design_single} to maximize the scanning range while still achieve comparable fill-factors to that of repeating patterns (derivations are provided in Supplementary Information). In the design rule, we search over ($f_x$, $f_y$) pairs around the resonance frequencies (in a close-to-far order) until we find a pair that falls in one of three ``good spatial coverage'' cases: Case1, where the scanning pattern repeats in $2T_{frame}$ time and a criterion in line~\ref{heuristic1} of design rule~\ref{alg:design_single} is met. Case2, where the scanning pattern repeats every $T_{frame}$ with $\phi_x = \phi_y = 0$ and Case3, where the scanning pattern repeats in $T_{frame}$ time with $\phi_x \neq \phi_y$. After the frequencies ($f_x$, $f_y$) are chosen, we determine the phases ($\phi_x$, $\phi_y$). The three ``good spatial coverage'' cases and the criterion in line~\ref{heuristic1} of design rule~\ref{alg:design_single} guarantee that the scanning trajectory does not repeat within $T_{frame}$. Mathematical proofs for the three ``good spatial coverage'' cases are provided in Supplementary Information. A very recent paper presented a design rule $|f_x\phi_y - f_y\phi_x|m = \pi/2$ to achieve a high fill-factor for repeating patterns~\cite{liss_design1}, which is similar to the phase selection rule in Case3. However, design rule~\ref{alg:design_single} is more complete and performs better under the physical constraints.

Figures~\ref{teaser}(b),(c) quantitatively show the dependence of fill-factor and scanning range on different settings $r \in [1,3]$ and $T_{frame} \in [6, 9]$. The figures compare the metrics of our proposed designs (blue, solid) and traditional repeating pattern designs~\cite{liss_design1} (black, dashed), with a fixed actuation amplitude of $1$ in all cases. The comparisons show that: (1) In most cases, the proposed designs have the same fill-factor as the repeating patterns, but larger scanning range. (2) When $T_{frame}$ is shorter, and when $m$ has more prime factors, the trade-off is generally less favorable. This is because with $f_x = k/4m$, the greatest common divider (GCD) of $k$ and $4m$ are usually larger than $4$ and do not fall in the three ``good spatial coverage'' cases in design rule~\ref{alg:design_single}. (3) Integer frequency ratios $r$ lead to worse trade-offs between fill-factor and scanning range. More discussions about this special case is provided in the discussion section. 
    
    \begin{algorithm}[t!]
    \caption{Design rule 1: unmodulated scanning patterns}\label{alg:design_single}
    \hspace*{\algorithmicindent} \textbf{Input} $f_x^r = r$, $f_y^r = 1$, $T_{frame} = m$\\
    \hspace*{\algorithmicindent} \textbf{Fixed} $f_y = 1$, $\phi_y = 0$\\
    \hspace*{\algorithmicindent} \textbf{Output} $f_x$, $\phi_x$\\
    \begin{algorithmic}[1]
    \State $k^{*} = \underset{k}{\operatorname{argmin}} |\frac{k}{4m} - r|$, initial searching space for k is all positive integer numbers.
    \While{TRUE}\Comment{Loop until find solution}
    \If{GCD($k^{*}$, $4m$) == 1}
    \Comment{GCD: the greatest common divider}
    \If{mod($k^{*}n, 4m$) $\neq \pm 1$, $n \in \{[m/2], [m/2] + 1, ... 3[m/2]\}$}\label{heuristic1}
    \State \Return $f_x = k^{*}/4m$, $\phi_x = 0$
    \Comment{Case1}\label{case1}
    \Else\Comment{Choose a sub-optimal $k$, continue loop}
    \State remove current $k^{*}$ from searching space
    \State $k^{*} = \underset{k}{\operatorname{argmin}} |\frac{k}{4m} - r|$
    \EndIf
    \EndIf
    
    \If{GCD($k^{*}$, $4m$) == 2}
    \State \Return $f_x = k^{*}/4m$, $\phi_x = 0$, 
    \Comment{Case2}\label{case2}
    \EndIf
    
    \If{GCD($k^{*}$, $4m$) == 4}
    \State \Return $f_x = k^{*}/4m$, $\phi_x =$ $\pi/(2m)$, 
    \Comment{Case3}\label{case3}
    
    \Else\Comment{Choose a sub-optimal $k$, continue loop}
    \State remove current $k^{*}$ from searching space
    \State $k^{*} = \underset{k}{\operatorname{argmin}} |\frac{k}{4m} - r|$ 
    \EndIf
    \EndWhile
    \end{algorithmic}
    \end{algorithm}
    
\subsection*{Modulated pattern design}
%\label{sec:optim_design}
    We further consider a more challenging operation of resonant scanners: Regions-of-Interest (RoI) focusing. In a LiDAR system, through-put of the 3D sensor is fixed, which makes adaptive spatial sampling beneficial for various applications~\cite{adaptive_sample1, adaptive_sample2, adaptive_sample3}. More specifically, given a user-defined RoI, we aim at sampling the RoI as densely as possible in all frames. This type of scanning is particularly challenging for resonant scanners and is beyond the capability of the unmodulated scanning patterns, so we propose to use modulated scanning patterns. Such patterns contain multiple frequency components within the resonance bandwidth. However, due to the higher degrees-of-freedom and the complexity of user-defined RoI, analytical design rules are inadequate, so to design modulated scanning patterns, we develop an optimization-based approach. The framework is task-driven because different imaging tasks have different Regions-of-Interest (RoI) for spatial sampling. We seek to improve, by optimized modulation of the parameters, the operation of the resonant scanner as characterized by Equation~\ref{eqn:motion1}. However, this model has continuous input parameters so we simplify the model through a Fourier expansion:
    \begin{equation}
    \label{eqn:motion2}
    \left\{ \begin{array}{l}
    x(t) = \sum\limits_{n = n_1}^{n_2} \alpha_n H_x(\frac{n}{Lm}) \textrm{cos}(2\pi\frac{n}{Lm} t) + \gamma_n H_x(\frac{n}{Lm}) \textrm{sin}(2\pi\frac{n}{Lm} t),\;\;\;\; \sqrt{\sum_n \alpha_n^2 + \gamma_n^2} <= 1\\
    y(t) = \sum\limits_{j = j_1}^{j_2} \beta_j H_y(\frac{j}{Lm}) \textrm{cos}(2\pi\frac{j}{Lm} t) + \delta_j H_y(\frac{j}{Lm}) \textrm{sin}(2\pi\frac{j}{Lm} t),\;\;\;\; \sqrt{\sum_j \beta_j^2 + \delta_j^2} <= 1\\
    \end{array} \right.
    \end{equation}
    
    where $H_x$ and $H_y$ are transfer function amplitudes. We ignore the phases of the transfer functions because they are included in the coefficients of the cosine and sine terms. $m$ specifies the frame time $T_{frame}$. $n_1$, $n_2$, $k_1$, $k_2$ defines the number of frequency components in optimization. We find that generally, $5$ frequency components give very good optimization results, and in most cases, $3$ frequency components are enough. $L$ is an integer that controls the spacing of the frequency components. Note that the amplitude constraints in Equation~\ref{eqn:motion2} are equivalent to bounding the root-mean-square ($\textrm{RMS}$) amplitudes of the actuation signals. This is a looser constraint than bounding the absolute actuation amplitude in unmodulated scanning, which can be expressed as $\sum_n \sqrt{\alpha_n^2 + \gamma_n^2} <= 1$, $\sum_n \sqrt{\beta_n^2 + \delta_n^2} <= 1$.
    
    From Equation~\ref{eqn:motion2}, we notice that the scanner motion is linearly determined by the parameter set $\{\alpha_n\}$, $\{\beta_k\}$, $\{\gamma_n\}$, $\{\delta_k\}$. Also, due to the band-pass characteristics of the transfer functions $H_x$, $H_y$, only frequency components close enough to resonant frequencies $f_x^r$, $f_y^r$ have significant impact on scanner motion. This allows efficient optimization of the parameter set. We further discretize time in Equation~\ref{eqn:motion2} to get the sampled scanning pattern $\textbf{x} \in \mathbb{R}^{N}$, $\textbf{y} \in \mathbb{R}^{N}$, where $N$ is the number of sampling points. The resonant frequencies $f_x^r, f_y^r$, frame time $T_{frame}$ and $N$ are chosen as hyper-parameters in the optimization.
    
\begin{figure}[t!]
    \begin{center}
    % \fbox{\rule{0pt}{3in} \rule{0.9\linewidth}{0pt}}
    \includegraphics[width=1.0 \linewidth]{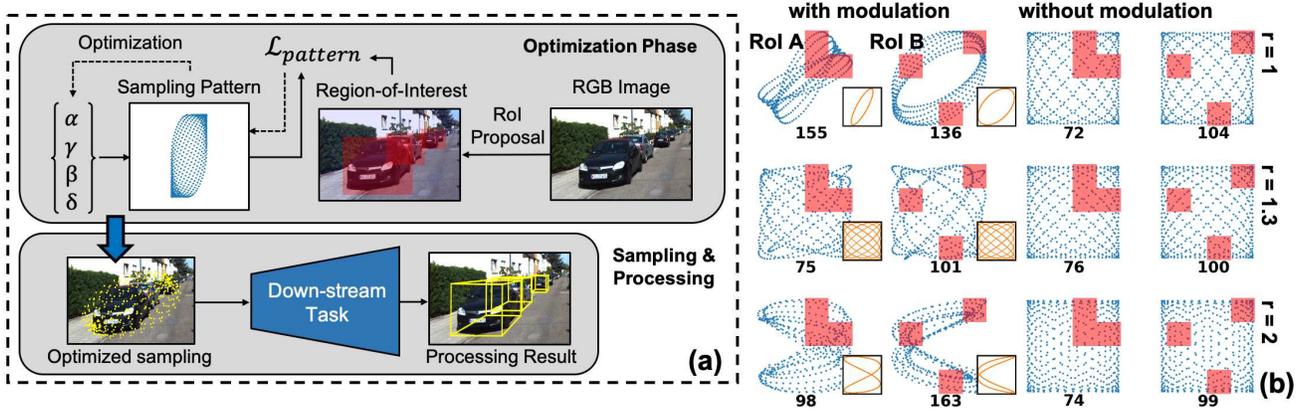}
    \end{center}
       \caption{(a) Schematic pipeline of the proposed optimization framework. It shapes the sampling pattern into task-specific (or even scene-specific) RoI-focused patterns through objective function $\mathcal{L}_{pattern}$. Here 3D object detection is used as a target task. (b) Optimization results for different RoI and resonant frequency ratios. Red rectangles show the specified RoI and black numbers under each pattern show the amount of sampling points within the RoI. With $r \sim 1$ and $r \sim 2$, the modulated scanning patterns have denser sampling in RoI, compared to the reference unmodulated scanning patterns. However, with $r \sim 1.3$, this RoI focusing improvement is not significant. At the lower-right corner of each modulated pattern, we also show the corresponding basic unmodulated pattern.}
    \label{pipeline}
    \end{figure}

    The optimization framework is shown in Figure~\ref{pipeline}(a). First, the parameter set is converted into a sampled scanning pattern $\textbf{x}$, $\textbf{y}$. For the specific task (in Figure~\ref{pipeline}(a), we use 3D object detection as an example), Regions-of-Interest (RoI) are proposed by a fast processing on 2D RGB image, or other heuristic rules and sensing results. The RoI is represented by a weight map $\textbf{W}$ and its values correspond to the importance of each regions in the FoV. With $\textbf{x}$, $\textbf{y}$ and $\textbf{W}$, we define the objective function $\mathcal{L}_{pattern}$ as:
    \begin{equation}
    \label{eqn:obj}
    \mathcal{L}_{pattern} = \sum_{i,j}^M\bar{W}_{i,j}[(x_{i} - \textbf{x}[n_{i,j}])^2 + (y_{j} - \textbf{y}[n_{i,j}])^2]
    \end{equation}
    The $[-1,1]\times[-1,1]$ FoV (normalized by the product of amplitudes with on-resonance actuation) is divided into $M\times M$ patches. For each patch $(i,j)$, we get the closest sampling point ($\textbf{x}[n_{i,j}]$, $\textbf{y}[n_{i,j}]$) to its center location ($x_{i}, y_{j}$) and calculate the distance between these two points. $\bar{W}_{i,j}$ indicates the importance of each patch and is defined as the average weight in patch $(i,j)$. Patches with larger average weights have a higher priority during optimization. Note that if the distance between patch $(i, j)$ and ($\textbf{x}[n_{i,j}]$, $\textbf{y}[n_{i,j}]$) is smaller than a threshold, this patch is considered as occupied and $\bar{W}_{i,j}$ is set to zero, regardless the weight value in this patch. From $\mathcal{L}_{pattern}$, gradient decent optimization~\cite{gradient_descent} is performed on the parameter set $\{\alpha_n\}$, $\{\beta_k\}$, $\{\gamma_n\}$, $\{\delta_k\}$. Once the optimization is done, spatial sampling can be conducted on a 3D scene, and a sparse point cloud is generated. The sampling is concentrated in the RoI, where most useful information is distributed, and the performance of down-stream tasks is improved. Note that this optimization need not to be done online (e.g. during scanner operation). For some tasks, optimized patterns for different scenes are very similar and thus the optimization process can be done off-line. An example is discussed in Supplementary Information.

     Figure~\ref{pipeline}(b) shows optimization results for two randomly-selected RoI (RoI A, B) and different resonant frequency ratios. We use $T_{frame} = 7$, total sampling point number $N = 500$ in all optimizations. $5$ frequency components are used in the optimization, with $f_x = \{6/7, 13/14, 1, 15/14, 8/7\}f_x^r$, $f_y = \{6/7, 13/14, 1, 15/14, 8/7\}f_y^r$. Similar results are achieved with $3$ frequency components $f_x = \{13/14, 1, 15/14\}f_x^r$, $f_y = \{13/14, 1, 15/14\}f_y^r$ and the effect of considering more frequency components (more than $5$) within the range of $[6/7, 8/7]f_x^r$, $[6/7, 8/7]f_y^r$ is not significant. The first and the second columns show optimization results with modulated scanning patterns. The third and the fourth columns show reference unmodulated scanning patterns. The modulated scanning patterns have bounded RMS actuation amplitude, as defined in Equation~\ref{eqn:motion2}, while the actuation amplitudes for unmodulated patterns are not bounded to better visualize the differences in RoI focused sampling. If the actuation amplitudes of unmodulated patterns are bounded, RoI on the edges and corners can't be reached in some cases. The comparisons show that: (1) With $r \sim 1$, $r \sim 2$, modulation and optimization lead to an improvement in sampling densities within the RoI, which are shown by the black number under each scanning pattern. (2) The RoI focusing improvement depends on the shape of the specified RoI. With $r \sim 1$, the RoI focusing is more successful for RoI A compared to RoI B, while it is the opposite with $r \sim 2$. (3) With $r \sim 1.3$, the RoI focusing improvement is very limited. The optimization results are only slightly improved compared to the unmodulated scanning patterns. 
     
     We provide a qualitative explanation for this dependence of RoI focusing improvement on resonant frequency ratio: A modulated scanning pattern ``dithers'' around a basic unmodulated scanning pattern, as shown in Figure~\ref{pipeline}(b) at the lower-right corner of each modulated pattern. This basic pattern corresponds to one pair of ($f_x$, $f_y$) in Equation~\ref{eqn:motion2} (it also needs to be in the resonance bandwidth). If the basic pattern has a short repeating period, it only traverse part of the scanning range. For example, with $f_x = 1, f_y = 1$, the scanning trajectory is a simple ellipse. Shape of the basic pattern is controlled by its amplitudes and phases in x and y-axis motion. When appropriate modulations are added, a small shift exists between the scanning trajectories in different repeating periods, and this leads to a focused sampling in the regions close to the basic pattern. However, if the basic pattern has a long repeating period and covers the scanning range uniformly, the modulated scanning patterns can't be focused onto a certain portion of FoV through optimization. For $r \sim 1$, $r \sim 2$, the repeating period is very short with $f_x = f_y$, $f_x = 2f_y$. However, for $r \sim 1.3$, there does not exist a ($f_x$, $f_y$) pair close enough to resonance while also has a short repeating period (for example, shorter than $t = 2$).

\begin{figure}[t!]
\begin{center}
% \fbox{\rule{0pt}{3in} \rule{0.9\linewidth}{0pt}}
\includegraphics[width=0.9 \linewidth]{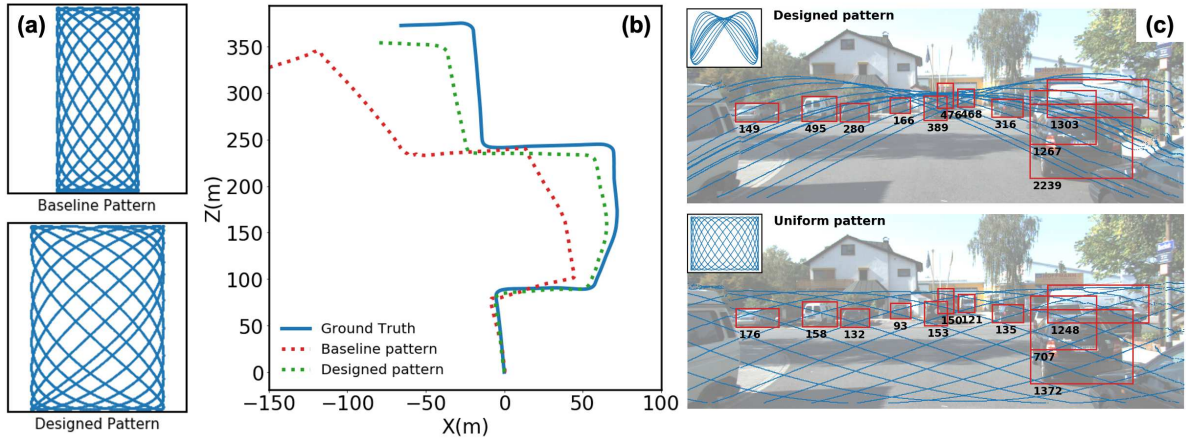}
\end{center}
   \caption{(a) Qualitative comparison of baseline and designed scanning patterns ($P_1$ and $P_2$ in Figure~\ref{teaser}(a)). Scanning range of the designed pattern is $1.6\times$ of that in baseline pattern, leading to a larger perspective field and more reliable feature extraction. (b) Trajectory estimations with optimized scanning pattern and baseline scanning pattern. (c) Object detection with optimized pattern and baseline pattern. Each red bounding box contains an object and the black number at bottom indicates number of sampling points contained in the bounding box. Due to the RoI-focusing improvement, optimized pattern contains $\sim 3\times$ more sampling points in bounding boxes. The sampling patterns are shown in blue dots.}
\label{odom_obj}
\end{figure}

\section*{Simulations}
%\label{sec:simu}
We evaluate the analytical design rule~\ref{alg:design_single} and the proposed optimization framework in simulated 3D environments~\cite{KITTI, NYUV2}. Because most 3D imaging datasets currently available are acquired with a raster-scanned or a flash LiDAR, we develop a point cloud generation tool that generates a point cloud corresponding to a resonant scanning pattern. Details of the dataset, implementation and more simulation results are provided in Supplementary Information.

\subsection*{LiDAR odometry with unmodulated scanning}
%\label{sec:odom}
LiDAR odometry algorithms estimate the trajectory of a moving agent during navigation. They extract feature points from a 3D point cloud acquired in each frame. By comparing the spatial positions of these feature points between successive frames, the position of the agent in a world coordinate can be estimated.

In this work, we consider LiDAR odometry with resonant scanning patterns on the KITTI dataset~\cite{KITTI}. We adapt a LiDAR odometry framework, named ``LOAM'', ~\cite{lidar1, livox-loam} into the resonant scanning scenario. For comparison, we use the example discussed in Figure~\ref{teaser}(a) with pattern $P_1$ as the baseline and pattern $P_2$ as the designed pattern. As shown in Figure~\ref{odom_obj}(a), Field-of-View (FoV) of $P_2$ is $\sim 1.6\times$ larger than that of $P_1$. This much larger spatial region gives us more feature points to be observed and processed, which leads to more reliable trajectory estimation (for details of the extracted feature points, please refer to Supplementary Information).

\subsection*{3D object detection with moduated scanning}
%\label{sec:detect}
Object detection is another task that is of great interest in 3D computer vision~\cite{pointnet, fpointnet}. The requirement it imposes on data collection is different from that in odometry. For each scene, important objects (e.g., cars, pedestrians) might concentrate in specific regions in the FoV. Therefore, a denser sampling in these Regions-of-Interest (RoI) is required. As an example, we use hyper-parameters $f^r_x = 1$, $f^r_y = 2$, $T_{frame} = 7$ and number of sampling points $N = 30000$. In Figure~\ref{odom_obj}(c), we show the RoI-focused scanning pattern (upper row). The pattern consists of three frequency components ($f_x = \{\frac{13}{14}, 1, \frac{15}{14}\}$, $f_y = \{\frac{27}{14}, 2, \frac{29}{14}\}$). The relative phases and amplitudes of the three components are optimized to be ($\phi_x = \{86^{\circ}, 178^{\circ}, 86^{\circ}\}$, $\phi_y = \{-96^{\circ}, 145^{\circ}, -96^{\circ}\}$, $A_x = \{0.22, 0.95, 0.22\}$, $A_y = \{0.28, 0.91, 0.28\}$). Using more frequency components in this special case won't generate significant improvements. When compared to the sampling pattern designed for uniform sampling (lower row), the RoI-focused pattern samples significantly more points ($\sim 3\times$) in regions that contain important objects (cars in this scene). This will largely facilitate the object detection process~\cite{fpointnet}. We do not conduct quantitative comparisons on object detection, due to the imperfectness in resonant-scanned point cloud generation. However, because of the positive relationship between sampling density and detection accuracy presented in previous literature~\cite{KITTI, fpointnet}, it is reasonable to expect an increase in accuracy when the optimized scanning pattern is used in real-world LiDAR system. 

Note that in this task we do not follow the $f^r_x = r$, $f^r_y = 1$ setting. This is because the dataset we experiment on contains only road scenes. Such a scene is more likely to be symmetric in the horizontal direction compared to the vertical direction. For example, cars are more likely to be on the left and right sides of a road, instead of on the up and down sides of a road. As mentioned above, the performance of RoI focusing depends on the RoI shape. When the axis of symmetry of the RoI shape aligns with that of the scanning pattern, performance is improved. Therefore, we make the scanning pattern also symmetric in horizontal direction by choosing $f^r_x = 1.0$, $f^r_y = 2.0$ instead of $f^r_x = 2.0$, $f^r_y = 1.0$. Optimization results with $f^r_x = 2.0$, $f^r_y = 1.0$ are also presented in Supplementary Information, where the performance is not as good as that in Figure~\ref{odom_obj}, but still beats the reference unmodulated scanning pattern.

\begin{figure}[t!]
\begin{center}
% \fbox{\rule{0pt}{3in} \rule{0.9\linewidth}{0pt}}
\includegraphics[width=0.8\linewidth]{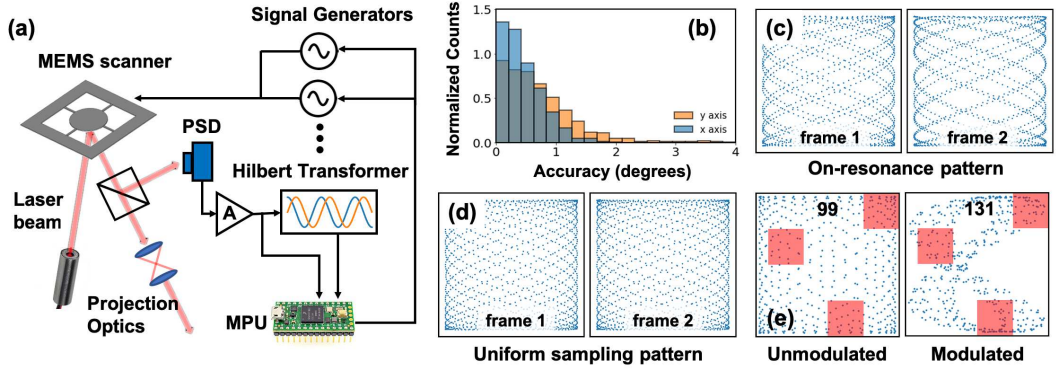}
\end{center}
   \caption{(a) Schematic of experimental set up for phase controlled resonant scanning. (b) Phase control accuracy of proposed hardware. (c) Recorded on-resonance scanning pattern, for two successive frames. (d) Recorded designed unmodulated sampling pattern, for two successive frames. (e) Recorded modulated sampling pattern with $r \sim 2.0$. Red rectangles are the Regions-of-Interest (RoI) and black numbers indicate amount of sampling points within RoI. Compared to the reference unmodulated scanning pattern, RoI sampling density is increased by $1.3\times$ with modulated scanning.}
\label{exp_sys}
\end{figure}

\section*{Experiments}
%\label{sec:exp}
We implement the designed scanning patterns (Figure~\ref{exp_sys}(a)) using a MEMS scanner~\cite{jwj} with resonant frequencies $f_x^r = 2660$Hz, $f_y^r = 1100$Hz, i.e., a resonant frequency ratio $r = 2.42$. The quality factors for the two axis are $Q_x \sim 30$ and $Q_y \sim 50$. Because of the high Q factor and associated low bandwidth, we actuate the y-axis with a single frequency, and restrict modulation to the x-axis. A high-gain amplifier is used to maintain the scanning range when we operate at more than FWHM (Full Width at Half Maximum) away from the resonance. We developed a wide-band phase detection and control system to eliminate the inherent phase uncertainty in MEMS scanners. This uncertainty originates from the environmental sensitivity (e.g. to temperature) of MEMS devices and the strong dependence of the phase on deviations of the resonant frequency~\cite{sandy_thesis, pll1, pll2, dual-tone}. With the control system, we achieve $\sim 1^{\circ}$ phase control accuracy, as shown in Figure~\ref{exp_sys}(b). To measure the accuracy, we detect the scanner phase at beginning of each frame and compare it to the required phase, over 10 minutes of scanner operation. This calibration is conducted with a high-speed oscilloscope not shown in Figure~\ref{exp_sys}(a). The accuracy can be improved with faster MPU or better position detection hardware. Phase stability with and without control are further discussed in the Supplementary Information.

\subsection*{Phase control in unmodulated scanning}
%\label{sec:single_exp}
We first demonstrate unmodulated scanning. During the experiments, the scanning patterns are recorded with a high-speed position sensor (PSD). We choose $T_{frame} = 6.4ms$, corresponding to $T_{frame} = 7$ in design rule~\ref{alg:design_single}. Scanning patterns with on-resonance actuation ($f_x = 2660$ Hz, $f_y = 1100$ Hz) and without phase control are shown in Figure~\ref{exp_sys}(c), for two successive frames. Most portions of the FoV are either over-sampled or under-sampled. Using our proposed design rule, the parameters are changed to $f_x = 2672$ Hz, $f_y = 1100$ Hz, $\phi_x = \pi/14$ and $\phi_y = 0$. The corresponding scanning patterns have much higher fill factor as shown in Figure~\ref{exp_sys}(d).

\subsection*{Phase control in modulated scanning}
%\label{sec:multi_exp}
To demonstrate modulated scanning, we drive the x-axis at three frequencies $f_x$, $13/14f_x$, $15/14f_x$ and drive the y-axis at a single frequency $f_y$. Phases of the three components in the x-axis scanning are monitored and controlled at the beginning of every 2 frames (when all three phases repeat). Resonant frequencies and frame time are set to be the same as that in the unmodulated scanning experiment. As discussed above, with $r = 2.42$, RoI focusing improvement is limited. Therefore, we go beyond the x-axis resonance bandwidth and select $f_x$ actuation frequency components around $2200$Hz while fix $f_y = 1100$Hz  to emulate a MEMS scanner with $r = 2.0$. We focus the scanning pattern to RoI B in Figure~\ref{pipeline}(b) for demonstration. Due to the high quality factor in y-axis, the degrees-of-freedom in optimization is reduced by $2\times$. However, RoI sampling density in modulated scanning pattern still increases by $1.3\times$ compared to the unmodulated scanning pattern, as shown in Figure~\ref{exp_sys}(e). The experimentally acquired modulated sampling pattern is resampled to $500$ sampling points per $T_{frame}$ for comparison with the sampling patterns in Figure~\ref{pipeline}.

\section*{Discussion}

It is important to note how performance depends on resonant frequency ratios for unmodulated scanning design rule~\ref{alg:design_single} and modulated scanning. Different ($f_x$, $f_y$) pairs generate unmodulated scanning patterns with different repeating periods. For any resonant frequency ratio $r$, pairs of ($f_x$, $f_y$) that correspond to long repeating period always exist in resonance bandwidth~\cite{hdhf1}. However, a pair of ($f_x$, $f_y$) that corresponds to short repeating period might not exist, as in the case of $r \sim 1.3$. Also, a pair of ($f_x$, $f_y$) that corresponds to repeating period $1/2T_{frame}$, $T_{frame}$ or $2T_{frame}$ do not always exist, as in the case of $r \sim 1$, $r \sim 2$. n the first situation, RoI focusing can't be achieved while in the second situation, uniform spatial sampling is difficult. In the experiments, we noticed that RoI focusing performs efficiently only with $r \sim 1$ or $r \sim 2$ while these are the worst cases in uniform spatial sampling, as shown in Figure~\ref{teaser}. This result suggests the special usage for resonant scanners with resonance frequency ratio $r \sim 1$, $r \sim 2$ in RoI focused sampling.

Although the proposed scanning pattern designs outperform the baselines, they have the following limitations. First, both designs are based on a moderate quality factor $Q$. If the quality factor is too high, neither the frequency selection rule in design rule~\ref{alg:design_single} nor the modulated scanning pattern designs produce good results. Only small deviations from resonance requires large actuation amplitudes, which is inconsistent with our bounded actuation setting. Second, the optimization problem in modulated scanning pattern design is non-convex. Therefore, our approach does not guarantee convergence to a global optimal. We also assume no cross-talk between x and y-axis scanner motions. This is consistent with the negligible cross talk we observe in our MEMS scanners~\cite{jwj}. If scanners with significant cross talk are employed, then the design rule for unmodulated patterns have to be changed to give good results. ROI focusing, on the other hand, does not need substantial changes to work with scanner that have cross talk. It is straightforward to contain the cross-talk in Equation~\ref{eqn:motion2} and use the optimization framework for both uniform and RoI-focused spatial sampling design.

\section*{Conclusion}
Spatial information acquisition is at the heart of many recent advances in the imaging and display industry. A fast and flexible spatial sampling solution will largely improve the robustness and consumer experience. In this paper, we propose resonant scanning pattern design and control schemes that improve the coverage, flexibility, and accuracy in fast spatial sampling. We propose an analytical design rule for uniform spatial sampling, and an optimization-based framework for flexible, Regions-of-Interest (RoI) focused spatial sampling. We also demonstrate the designed scanning patterns in an experimental prototype that applies wide-band control on scanner motion. The proposed methods enable resonant-scanner LiDAR with a high frame-rate $\sim 100$Hz. When integrated with high-speed point cloud processing algorithms, such systems can be utilized in applications across disciplines, including navigation, robotics, and augmented reality. 

% \section*{Conclusion}

\section*{Methods}
\subsection*{Phase control experimental setup}
As shown in Figure~\ref{exp_sys}(a), the MEMS scanner is actuated with signal generators (SIGLENT SDG2000X) controlled by external phase modulation signals. The motion of MEMS is detected with a high-speed position sensor (ON-TRAK OT-301). This motion signal is fed into an analog wide-band Hilbert transformer board for 90 degrees phase shift. Both motion signals $x(t)$, $y(t)$ and the 90degrees phase shifted signals $\bar{x}(t)$, $\bar{y}(t)$ are sampled with an MPU chip (PJRC Teensy3.6). In practice, the Hilbert transformer applies a frequency dependent phase shift on both output signals while the relative phase between these two outputs is fixed to be $\pi/2$. We conduct calibrations to remove the phase offset and will ignore it in the following sections. For more details, please refer to Supplementary Information. A fast processing algorithm is performed on the two signals to get the phase and a feed-back signal is generated to the external modulation port of signal generators. Calibrations for each components used in experiment are provided in Supplementary Information.

\subsection*{Phase calculation process}
Phase calculations are simple for the unmodulated actuation case. After collecting $x(t)$ and $\bar{x}(t)$ at the beginning of each frame, a fast arctangent calculation~\cite{fast_atan} is performed to get the phase. The detection process takes $\sim 15$us. 

For modulated actuation, phase detection and control is more complicated, because $x(t)$, $y(t)$ and $\bar{x}(t)$, $\bar{y}(t)$ contain multiple frequency components. In this paper, we constrain ourselves to a comparatively simple situation: $x$ axis actuation contains three frequency components and $y$ axis contains only single frequency component. Similar method can be extended to a more general case. We express the scanner motion in the x-axis as:
\begin{equation}
\label{eqn:exp_pdetect1}
\left\{ \begin{array}{l}
x(t) = \alpha_0 cos(\omega_x^0t + \phi_x^0) + \alpha_1 cos(\omega_x^1t + \phi_x^1) + \alpha_2 cos(\omega_x^2t + \phi_x^2)\\
\bar{x}(t) = \alpha_0 sin(\omega_x^0t + \phi_x^0) + \alpha_1 sin(\omega_x^1t + \phi_x^1) + \alpha_2 sin(\omega_x^2t + \phi_x^2)
\end{array} \right.
\end{equation}

There are three phases $\phi_x^i, i = 0,1,2$. We detect at both beginning of each frame and at the center of each frame to get six equations:
\begin{equation}
\label{eqn:exp_pdetect2}
\left\{ \begin{array}{l}
x(0) = \alpha_0 cos(\phi_x^0) + \alpha_1 cos(\phi_x^1) + \alpha_2 cos(\phi_x^2), \\ x(T_{frame}/2) = \alpha_0 cos(\omega_0T_{frame}/2 + \phi_x^0) + \alpha_1 cos(\omega_1T_{frame}/2 +  \phi_x^1) + \alpha_2 cos(\omega_2T_{frame}/2 +  \phi_x^2), \\
x(T_{frame}) = \alpha_0 cos(\omega_0T_{frame} + \phi_x^0) + \alpha_1 cos(\omega_1T_{frame} +  \phi_x^1) + \alpha_2 cos(\omega_2T_{frame} +  \phi_x^2), \\

\bar{x}(0) = \alpha_0 sin(\phi_x^0) + \alpha_1 sin(\phi_x^1) + \alpha_2 sin(\phi_x^2), \\ \bar{x}(T_{frame}/2) = \alpha_0 sin(\omega_0T_{frame}/2 + \phi_x^0) + \alpha_1 sin(\omega_1T_{frame}/2 +  \phi_x^1) + \alpha_2 sin(\omega_2T_{frame}/2 +  \phi_x^2), \\
\bar{x}(T_{frame}) = \alpha_0 sin(\omega_0T_{frame} + \phi_x^0) + \alpha_1 sin(\omega_1T_{frame} +  \phi_x^1) + \alpha_2 sin(\omega_2T_{frame} +  \phi_x^2), \\

\end{array} \right.
\end{equation}
Equation~\ref{eqn:exp_pdetect2} is linear in $\{ \alpha_i cos(\phi_x^i), \alpha_i sin(\phi_x^i) \}, i = 1,2,3$, a fast matrix multiplication is used to solve them. Then we apply the fast arctangent calculation on each $(cos, sin)$ pair separately to get the phases. The whole data acquisition and processing takes $\sim 40$us for three frequency components. 

\bibliography{sample}

\section*{Author contributions statement}
Z. Sun conducted the simulation and experimental work. R. Quan conducted circuit design and provided critical suggestions. O. Solgaard supervised the project. All authors reviewed the manuscript.     

\section*{Competing interests}
The author(s) declare no competing interests.

\section*{Data availability}
All data used in plotting the figures (including figures in Supplementary Information) are available at
\newline
\url{https://drive.google.com/drive/folders/1I5_auWKR-UVEHSugAnWAarbaphlW4iKL?usp=sharing}

\newpage
\section*{Spplementary Information}
\appendix

\section*{Unmodulated scanning pattern design}
\subsection*{Concepts of ``repeating pattern'' and ``frame time''}
We note that the terms ``repeating pattern'' and ``frame time'' are used in previous literature~\cite{high_fps1, high_fps2} while their definitions are different from those used in this paper. To avoid confusion, here we give a detailed comparison between the concepts:

In previous literature~\cite{high_fps1, high_fps2}, a ``non-repeating pattern'' is defined as a resonant scanning pattern with irrational scanning frequency ratio and never repeats. Any scanning pattern that repeats in limited time is denoted as a ``repeating pattern''. Accordingly, the ``frame time'' is defined as the repeating period of a resonant scanning pattern.

In this paper,we focus on optimization of scanning pattern in a short, given time $T_{frame}$. We denote $T_{frame}$ as ``frame time'', while it does not necessarily equal the repeating period of scanning pattern. The ``frame'' here refers to a time period in which one block of data is collected. This data block is processed individually and is not combined with data from other ``frames''. The ``frame time'' is a manually given input in the proposed design rule. Practically, it can be the safety response time of an autonomous vehicle and 
is specified by a higher-level system requirement. The concurrent research~\cite{liss_design1} shared similar concepts with us. However, they only considered scanning patterns that repeat in each ``frame time'' (denoted as ``repeating pattern'' in this paper) and ignore the physical constraint of maximum actuation amplitude in their design. The proposed design rule expands the design space to include patterns that does not repeat in each ``frame time'' (denoted as ``non-repeating pattern'' in this paper). Therefore, it is more complete and achieves better performance.

\subsection*{Derivation of unmodulated scanning pattern design rule}
%\label{si:derive_design}
The intuitive starting point of proposed unmodulated scanning pattern design rule is:

\textit{The scanning trajectory can't \textbf{repeat (or almost repeat)} itself at \textbf{middle} of a frame.} 
\newline
\newline
\noindent
As long as this criteria is met, the scanning pattern within $T_{frame} = m$ has a good fill-factor. Without loss of generality, we make the same assumption as in the main text (since we focus on unmodulated scanning pattern design in this subsection, we ignore the time-dependence of amplitudes $A_x$, $A_y$ and phases $\phi_x$, $\phi_y$):
\begin{equation}
    \left\{ \begin{array}{l}
    x(t) = A_x cos(2\pi f_xt + \phi_x), \;\; p_x(t) \equiv 2\pi f_xt + \phi_x\\
    y(t) = A_y cos(2\pi f_yt + \phi_y), \;\; p_y(t) \equiv 2\pi f_yt + \phi_y\\
    f_x = p/q, \;\; p, q ~\textrm{coprime}\\
    f_y = 1,\;\;\;\; \phi_y = 0
    \end{array} \right.
\end{equation}
Where $f_x$, $f_y$, $\phi_x$, $\phi_y$ are the frequencies and phases for two scanning axis. We define two new variables, $p_x(t)$, $p_y(t)$ as the instance phases, for simplicity. We define a set of $t$ such that $\textrm{mod}[p_y(t), \pi] = 0$. We denote this set of $t$ as $T_{nodes}$. With the above assumptions, $T_{nodes} = \{n/2\}, n \in \mathbb{Z}$. When $\textrm{mod}[p_y(t), \pi] = 0$ and $\textrm{mod}[p_x(t), \pi] = 0$ both hold, the scanning trajectory begins to repeat itself. If $t$ does not coincide with the beginning or end of a frame, the fill-factor of scanning pattern reduces significantly. Also, with $t \in T_{nodes}$, if $|\textrm{mod}[p_x(t), \pi]|$ is small, the scanning pattern also follows a trajectory that is close to self-repeating. To provide an optimal scanning pattern, we want $|\textrm{mod}[p_x(t), \pi]|$ to be as large as possible for $\forall t \in T_{nodes}$, except when $t$ is at the beginning or end of a frame. Starting from this observation, we list several conclusions that leads to the three cases in design rule 1:

\begin{itemize}
    \item First, repeating period of the scanning pattern (equals to $q$) must be synchronized with the frames. That is, $q = km, k \in \mathbb{Z}^+$. Otherwise, no matter how phases are chosen, the scanning pattern repeats its trajectory at least within some frames. This does not meet the criteria.
    
    \item \textbf{With $q = m$, which corresponds to Case3 in design rule 1}, selecting the phase $\phi_x$ is important. With $t \in T_{nodes} = \{n/2\}, n \in \mathbb{Z}$, $\rightarrow p_x(t) = np\pi/m + \phi_x$. Since $p, m$ coprime, $p_x(t)$ value traverses the set $\{k\pi/m + \phi_x\}, k \in \{0,1,2,...m-1\}$. Therefore, the minimum value of $|\textrm{mod}[p_x(t), \pi]|$ is either $|\phi_x - \pi/m|$ or $\phi_x$. If either one of these two values is small, the pattern is close to self-repeating. To make both values sufficiently large, we choose $\phi_x = \pi/(2m)$, such that $|\phi_x - \pi/m|$ also equals to $\pi/(2m)$.
    
    \item \textbf{$q = 2m$ corresponds to Case2 in design rule 1}. Similar to the above derivations, with With $t = n/2, n \in \{0,1,2,...,2m-1\}$, the minimum value of $|\textrm{mod}[p_x(t), \pi]|, t \in T_{nodes}$ is either $|\phi_x - \pi/(2m)|$ or $\phi_x$. Therefore, we choose $\phi_x = 0$, such that the minimum value of $|\textrm{mod}[p_x(t), \pi]|$ is only achieved at $t = 0$, the beginning of a frame.
    
    \item $q > 2m$, similar to the above derivations, with $t = n/2, n \in \{0,1,2,...,2m-1\}$, the minimum value of $|\textrm{mod}[p_x(t), \pi]|, t \in T_{nodes}$ is either $|\phi_x - \pi/q|$ or $\phi_x$. No matter how we choose $\phi_x$, the minimum value of $|\textrm{mod}[p_x(t), \pi]|$ won't be larger than $\pi/q$. Since $q > 2m$, usually fill-factor of the scanning pattern reduces significantly.
    
    \item \textbf{Heuristically, we find an exception with $q = 4m$, which corresponds to Case1 in design rule 1}. Although the minimum value of $|\textrm{mod}[p_x(t), \pi]|, t \in T_{nodes}$ equals to $\pi/(4m)$, it only happens once and at any other $t \in T_{nodes}$,  $|\textrm{mod}[p_x(t), \pi]| \geq \pi/(2m)$. If  $|\textrm{mod}[p_x(t), \pi]| = \pi/(4m)$ only happens close to beginning or end of frames, the fill-factor reduction can be ignored, as in the example discussed in main text (Figure 1(a), Pattern 2). Therefore, we add an additional check for Case1:
    \begin{equation}
    \begin{array}{l}
    \forall t \in T_{nodes}, |\textrm{mod}[t, m]| > [m/2],\;\; \textrm{(These $t$ values are not close enough to beginning or end of frames)}, \\
    |\textrm{mod}[p_x(t), \pi]| \neq \pi/(4m)
    \end{array}
    \end{equation}
    
    \item \textbf{We also give an explanation on why $q = 3m$ and $q = 5m$ are not selected}. Similar to Case3, when $q = 3m$, the minimum value of $|\textrm{mod}[p_x(t), \pi]|, t \in T_{nodes}$ equals to $\pi/(3m)$. It only happens once and at any other $t \in T_{nodes}$,  $|\textrm{mod}[p_x(t), \pi]| \geq 2\pi/(3m)$. If $|\textrm{mod}[p_x(t), \pi]| = \pi/(3m)$ only happens close to beginning or end of frames, the fill-factor reduction should be tolerable. However, suppose $t_1 \in T_{nodes}$, $|\textrm{mod}[p_x(t_1), \pi]| = \pi/(3m)$, we have $t_1 + 3m/2 \in T_{nodes}$, $|\textrm{mod}[p_x(t_1 + 3m/2), \pi]| = |\textrm{mod}[p_x(t_1) + p\pi, \pi]| = \pi/(3m)$. The separation between  $t_1$ and $t_1 + 3m/2$ is $3/2$ of a frame time $m$. Therefore, it is impossible for both $t_1$ and $t_1 + 3m/2$ to be close to beginning or end of frames, which means there are always some frames with low fill-factor.
    
    Similar explanation is valid for $q = 5m$.
    
    \item With $q \geq 6m$, the fill-factor reduction is worse than the case $q = 4m$. This is because even if the minimum value of $|\textrm{mod}[p_x(t), \pi]|$ is achieved close to beginning or end of frames, part of the $k$th minimum values, $k \geq 2$ are also small enough to significantly influence the fill-factor. \textbf{Therefore, we do not include the cases with $q > 4m$ in design rule 1}.

\end{itemize}

\subsection*{Fill-factor computation for unmodulated scanning patterns}
To quantitatively calculate the fill-factor for a scanning pattern, we first scale a scanning pattern into $[-1,1] \times [-1,1]$ range, because fill-factor should not depends on scanning range (size of scanning pattern). Then we sample $1000$ points from this scaled pattern, with equal time interval. We divide the $[-1,1] \times [-1,1]$ range into $128 \times 128$ patches. For each patch, we search for a minimum distance from its center to the set of sampling points. Finally, we take the maximum value among these minimum distances as $R_{max}$, radius of the largest inscribed circle.

\subsection*{Phase error tolerance in unmodulated scanning pattern designs}
\begin{figure}[t!]
\begin{center}
% \fbox{\rule{0pt}{3in} \rule{0.9\linewidth}{0pt}}
\includegraphics[width=1.0\linewidth]{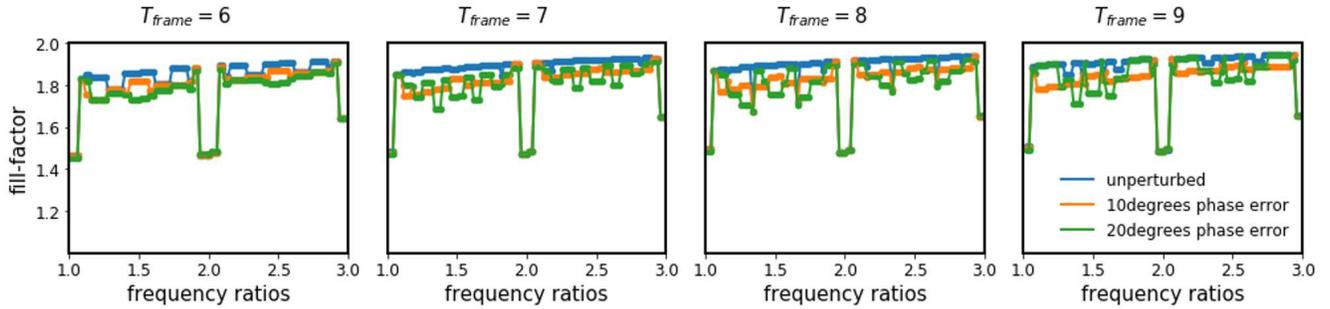}
\end{center}
   \caption{Analysis on phase error tolerance in the unmodulated scanning pattern design. A phase error of $10^{\circ}$/$20^{\circ}$ are added to $\phi_x$ when generating the scanning pattern. Compared to the non-perturbed patterns, this phase error results in a degradation in sampling efficiency.}
\label{appendix_tolerance}
\end{figure}

%\label{si:tolerance}
Small uncertainties in phases are unavoidable in real-world system, even with well-designed controls. Here we analyze the phase tolerance of the proposed unmodulated scanning design rule. In Figure~\ref{appendix_tolerance}, we plot the fill-factor vs. resonant frequency ratio $r \in [1,3]$, with a small phase shift $\delta \phi_x$ added to the designed phase $\phi_x$. As can be seen, with $\delta \phi_x$,
sampling efficiency significantly reduces, but is still at a reasonable value. Also, we noticed that the degradation in sampling efficiency does not grow with the phase error. This is due to the periodic nature of resonant scanning patterns.

\section*{Simulation details}
%\label{si:detail_simu}

We will provide all scripts used in simulations on publication:
\newline
\url{https://github.com/zhsun0357/Resonant-Scanned-LiDAR}
\subsection*{Dataset generation}
We generate the resonant-scanned point cloud from KITTI dataset~\cite{KITTI} and NYUV2 dataset~\cite{NYUV2} for scanning pattern evaluations. 
\noindent
For KITTI dataset, we first generate a dense depth map with depth inpainting algorithm~\cite{self_s2d}. We interpret $\textbf{x}$, $\textbf{y}$ in sampling pattern as the normalized yaw angle $\mathbf{\phi}$ and pitch angle $\mathbf{\theta}$. With the calibrated transformation matrix between camera coordinate and LiDAR coordinate, we are able to sample a 3D point cloud in LiDAR coordinate from the dense depth map with $\mathbf{\phi}$ and $\mathbf{\theta}$. There are two major limitations in this point cloud generation process. First, we crop the $360^{\circ}$ scanned point cloud in KITTI dataset into an horizontal angle (yaw angle $\phi$) range $\sim \pm 40^{\circ}$, corresponding to the camera FoV. This is because we need the RGB image as reference in depth inpainting~\cite{self_s2d} while naive depth inpainting without reference RGB image results in large error. Second, due to the pixelization and imperfect estimation in the dense depth map, the resulted resonant-scanned point cloud is geometrically distorted compared to raw point cloud from KITTI dataset. An example of the generated point cloud from KITTI dataset is shown in Figure~\ref{appendix_dataset}(a), in bird's-eye view. We crop out a local patch in the generated point cloud and compare it to the corresponding reference point cloud from raster-scanned, raw KITTI data. Although the surfaces (e.g., buildings, cars) are generally maintained, small geometric errors can be noticed. This distortion leads to difficulties when implementing object detection on sampled point cloud, because detection results on resonant-scanned point cloud can't be compared with ground truth for raw point cloud. Therefore, we do not show the quantitative comparison on object detection accuracy in the paper. Nevertheless, through the proposed task-driven optimization, enhancement in sampling density in the Regions-of-Interest is evident. Due to the positive relationship between sampling density and detection accuracy presented in previous literature~\cite{KITTI, fpointnet}, it is reasonable to expect an increase in accuracy when the optimized scanning pattern is used in real-world LiDAR system.

\noindent
For NYUV2 dataset, since a dense depth map is provided, straight-forward sampling on this dense depth map is conducted. Sampling pattern $\textbf{x}$ and $\textbf{y}$ are directly converted into pixel coordinates in the dense depth map.

\subsection*{Implementation}
\begin{figure}[t!]
\begin{center}
% \fbox{\rule{0pt}{3in} \rule{0.9\linewidth}{0pt}}
\includegraphics[width=1.0\linewidth]{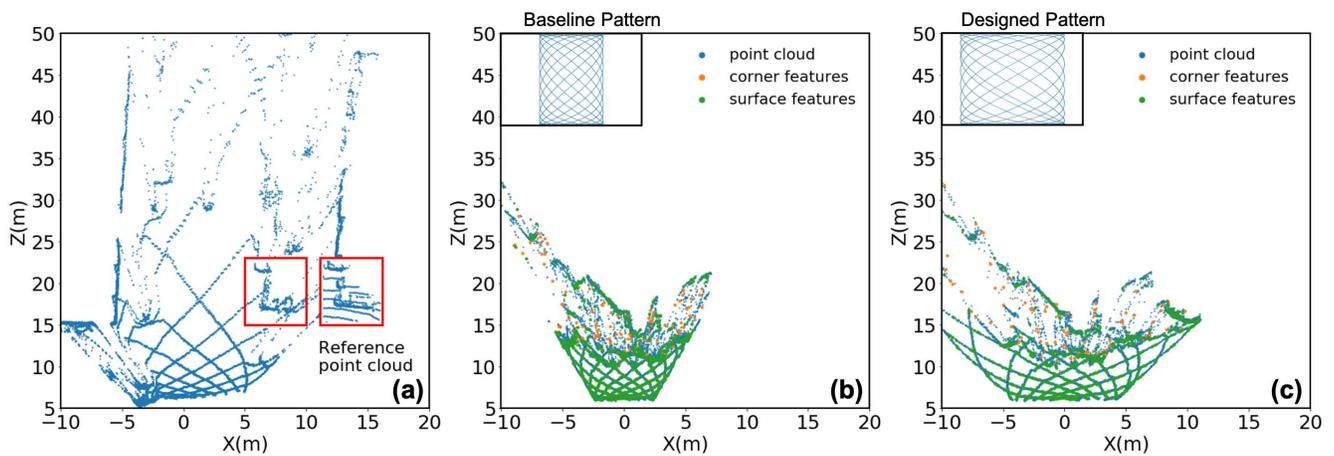}
\end{center}
   \caption{(a) An example generated resonant-scanned point cloud (in bird's-eye view). Note the distortions compared to corresponding raster-scanned ``reference point cloud'' in KITTI dataset. (b)(c) Feature points extraction (for both surface and corner features) with baseline/optimized sampling patterns in bird's-eye view. Blue points correspond to the full point cloud, orange and green points correspond to corner and surface feature points separately.}
\label{appendix_dataset}
\end{figure}

For LiDAR odometry task, we adapt the framework ``LOAM'' from previous literature~\cite{lidar1}. The trajectory estimation process can be divided into three steps: 1. Divide LiDAR point cloud into multiple scan sections 2. Corner/Surface feature extraction from each section and 3. Trajectory estimation with the extracted features. To better estimate the quality of data collected in a single frame, we do not contain a global mapping step in the algorithm. In Figure~\ref{appendix_dataset}(b)(c), we show extracted feature points with baseline and designed patterns. Two types of feature points are extracted: corner features that provide in-line geometric constraints and surface features that provide in-plane constraints. A larger Field-of-view (FoV) of designed pattern contains more high-quality feature points and thus makes the odometry estimation more reliable.

\begin{figure}[t!]
\begin{center}
% \fbox{\rule{0pt}{3in} \rule{0.9\linewidth}{0pt}}
\includegraphics[width=1.0\linewidth]{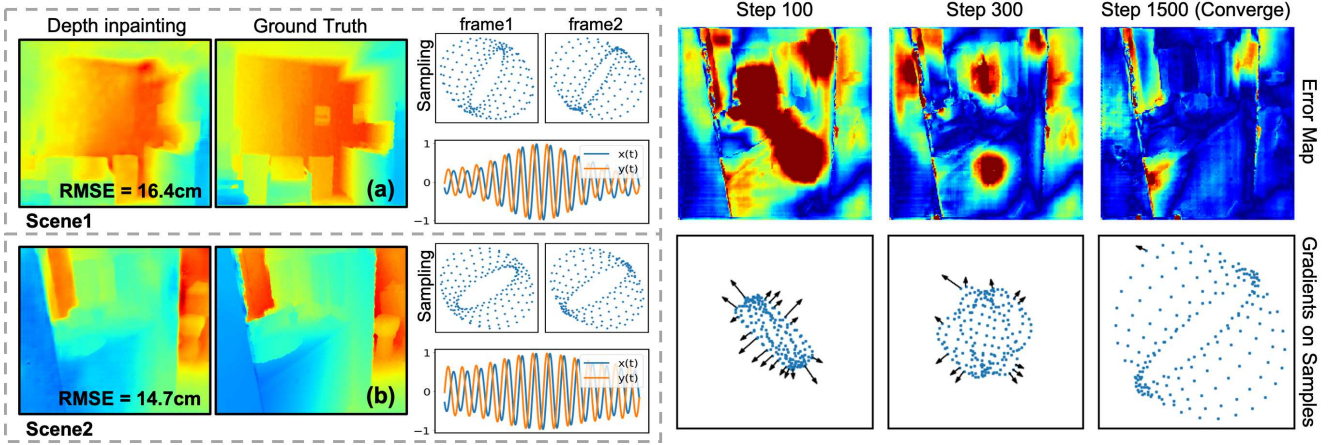}
\end{center}
   \caption{(a), (b) For two scenes in NYUV2 dataset, we show the optimized sampling pattern with $N = 200$ sampling points. We also show the modulated $x(t)$, $y(t)$ with optimized parameter set. On the left part of (a), (b), depth inpainting results with optimal sampling patterns are compared with ground truth depth map. Root-mean-square-errors (RMSE) are listed for each scene. (c) Weight (error) map and gradients on sampling pattern at different optimization steps. Amplitudes and signs of gradients are indicated by the lengths and directions of black arrows.}
\label{appendix_dcomp}
\end{figure}

Apart from the example of scanning pattern optimization in object detection (shown in main text), we demonstrate the mechanism of the proposed optimization framework with another computer vision task: depth inpainting. Depth inpainting involves generating a dense depth map from a sparsely sampled point cloud (and optionally, a reference RGB image). In this work, we adapt a state-of-the-art depth inpainting framework ~\cite{adaptive_sample1} for task-driven scanning pattern optimization. We use multiple scenes in NYUV2 dataset~\cite{NYUV2} to get optimized scanning patterns. Regions-of-Interest weight is estimated by the relative absolute error at each pixel during training. Hyper-parameters $f_x^r = 1$, $f_y^r = 1$, $T_{frame} = 7$ and number of sampling points $N = 200$ are used. 

As shown in Figure~\ref{appendix_dcomp}(a),(b), it turns out that optimizations with different scenes in the dataset converge to very similar optimal scanning patterns. This is due to the fact that in depth inpainting, each position in the whole FoV is of almost equivalent importance. Therefore, instead of an RoI-focused sampling, a better design for this task might be directly using the unmodulated design. As mentioned in the scanner motion model, in task-driven optimization, the constraint on actuation amplitude is loosened. For the example shown in Figure~\ref{appendix_dcomp}, optimized scanning pattern requires peak actuation amplitude $\sim 2.0$. To get comparable performance, unmodulated scanning pattern only requires an actuation signal with peak amplitude $\sim 1.3$.

The depth inpainting framework proposed by Bergman et al.~\cite{adaptive_sample1} consists of a rough bilateral filter stage~\cite{bsolver} and a refinement stage. The model also contains a monocular depth estimator~\cite{mono} to assist the inpainting task. The whole inpainting algorithm is accomplished by an end-to-end convolutional neural network (CNN), where the bilateral filter is also approximated by an CNN model. Instead of that, we use the original bilateral filter in the pipeline, and directly use the error-map from this bilateral filter stage for optimization.
The advantage of the proposed optimization framework, compared with that in previous literature~\cite{adaptive_sample1} is: RoI information across the whole FoV has impact on the scanning pattern updating, even when the non-optimized sampling region is small. Examples of the gradients on sampling pattern at different optimization steps are shown in Figure~\ref{appendix_dcomp}. It can be seen that at each step, gradients on sampling points are ``streching'' the sampling pattern to uncovered regions. Note that in an early step (step 100), the inpainting result is rough and RoI-weight (error) is even larger in the sampled region. This might be due to the fact that depth reconstruction in these initially sampled regions is coincidentally more difficult. However, occupied patches are set with zero weights in objective function. Therefore, despite this RoI-weight distribution, the optimization framework still managed to ``expand'' the sampling pattern instead of ``trapping'' it in a local region with large error.

\begin{figure}[t!]
\begin{center}
% \fbox{\rule{0pt}{3in} \rule{0.9\linewidth}{0pt}}
\includegraphics[width=1.0\linewidth]{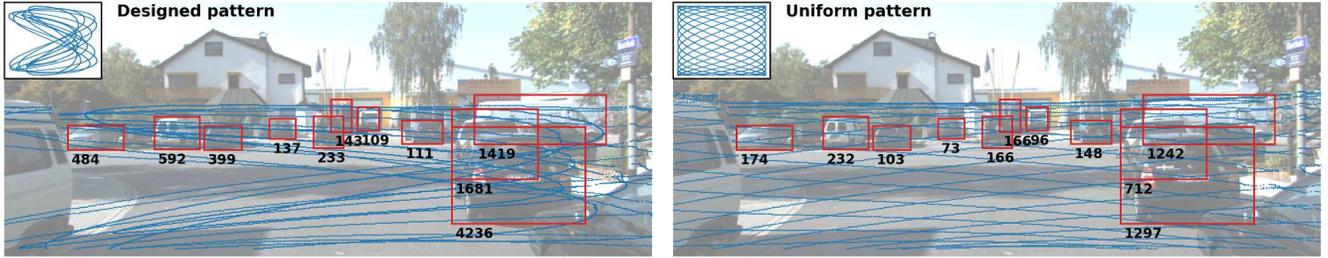}
\end{center}
   \caption{RoI focusing optimization result for 3D object detection with $f_x = 2.0$, $f_y = 1.0$.}
\label{appendix_obj21}
\end{figure}

We also show the RoI focusing optimization result for 3D object detection with $f_x = 2.0$, $f_y = 1.0$ in Figure~\ref{appendix_obj21}. Although the axis of symmetry of the scene does not align with that of the scanning pattern, modulated scanning pattern still out-performs the reference unmodulated scanning pattern significantly.

\begin{figure}[t!]
\begin{center}
% \fbox{\rule{0pt}{3in} \rule{0.9\linewidth}{0pt}}
\includegraphics[width=0.5\linewidth]{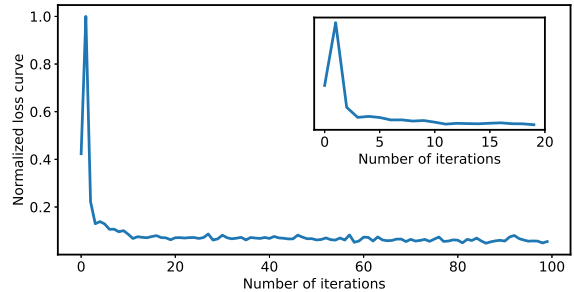}
\end{center}
   \caption{A typical convergence curve in the modulated scanning pattern optimization. The inset is a zoom-in view for iteration steps less than 20. Usually partial convergence is achieved within 10 to 20 iteration steps. With better initialization strategy, the convergence steps can be reduced to $< 5$ (not shown in the figure)}
\label{num_iter}
\end{figure}

\subsection*{Considerations in dynamic scenes}
In the main text, we focus on optimizing scanning pattern in each data collection period $T_{frame}$ and assume a static environment. In real-world LiDAR applications, adjusting scanning pattern design according to dynamic scenes is also an important functionality. Here we briefly analyze how to adapt the proposed designs to this scenario.

In unmodulated scanning pattern design, the goal is an optimal uniform spatial sampling. With a fixed resonant frequency ratio $r$ and a fixed frame time $T_{frame}$, the optimal scanning pattern is uniquely given by design rule 1 and is not related to the scene. The designed unmodulated scanning patterns repeat (Case2,3 in design rule 1) or almost repeat (Case1 in design rule 1) in each $T_{frame}$, regardless the changes in the scene. This is the same case as in most LiDAR systems and we demonstrate its effectiveness in the LiDAR odometry task.

In modulated scanning pattern design, the goal is RoI-focusing and in many cases, the optimized scanning pattern is preferred to change with the scene. When the RoI changes, to make the transition from a previously designed scanning pattern to an updated one, time delay is inevitable. In the proposed scanning pattern design and control framework, this time delay majorly consists of MEMS device response time and scanning pattern design computation time. The MEMS response time is roughly given by $t_{MEMS} = Q/\pi f^r$ under the assumption of ideal harmonic oscillator~\cite{impulse} ($Q$ is the quality factor of the MEMS device, $f^r$ is the MEMS resonant frequency). With $Q = 20$, $f^r = 1$kHz, $t_{MEMS} \sim 6$ms.

Computation time for modulated scanning pattern design depends on multiple factors: complexity of RoI weight map ($W$ in main text, Equation 4), number of iteration steps and the initialization strategy. The weight map can take float values as in the depth completion task (Figure~\ref{appendix_dcomp}), or binary values as in the 3D object detection task discussed in the main text (Figure 3(c)). This choice of RoI weight map parameters depends on the format of input RoI. In 3D object detection task, the RoIs are given as the 2D object bounding boxes estimations. Therefore, binary RoI weight map is sufficient to represent these simple geometric shapes.

With binary RoI weight map, the convergence of optimization framework is much faster and each optimization step size can be larger, compared to that shown in Figure~\ref{appendix_dcomp}. When started from a random initialization, the algorithm usually takes 10 to 20 iteration steps to achieve convergence. Figure~\ref{num_iter}(a) shows a typical convergence curve. We implement the optimization algorithm with PyTorch (for automatic gradient descent) and run it on Intel Core i7 CPU (Macbook Pro). Each iteration takes $\sim$ 2.5ms and the total optimization time is $\sim$ 25-50ms. We expect better implementation and more powerful hardware to increase the speed of single optimization iteration.

Although the pattern transition time delay roughly satisfies real-time operation requirement ($30$ FPS), we can further reduce the time delay for higher speed systems. One of the key insight is that real-world scenes change smoothly. Therefore, a Kalman filter-type prediction algorithm can be utilized to compensate for the time delay. Similar approaches are widely applied in LiDAR data stream processing pipelines~\cite{mot1, mot2}. Also, since regions-of-interest in successive frames overlap significantly, optimized patterns should also be similar. Therefore, we can use the optimization result in the previous frame as the initialization in current frame optimization. Preliminary experiments show that usually less than 5 iteration steps is needed for convergence, in contrast to $\sim 20$ iterations when the optimization is randomly initialized. Detailed discussions on these algorithm improvements are out of the scope of the current paper and we leave them for future work.

\section*{Experimental details}

\subsection*{Calibrations on phase control system}
\begin{figure}[t!]
\begin{center}
% \fbox{\rule{0pt}{3in} \rule{0.9\linewidth}{0pt}}
\includegraphics[width=1.0\linewidth]{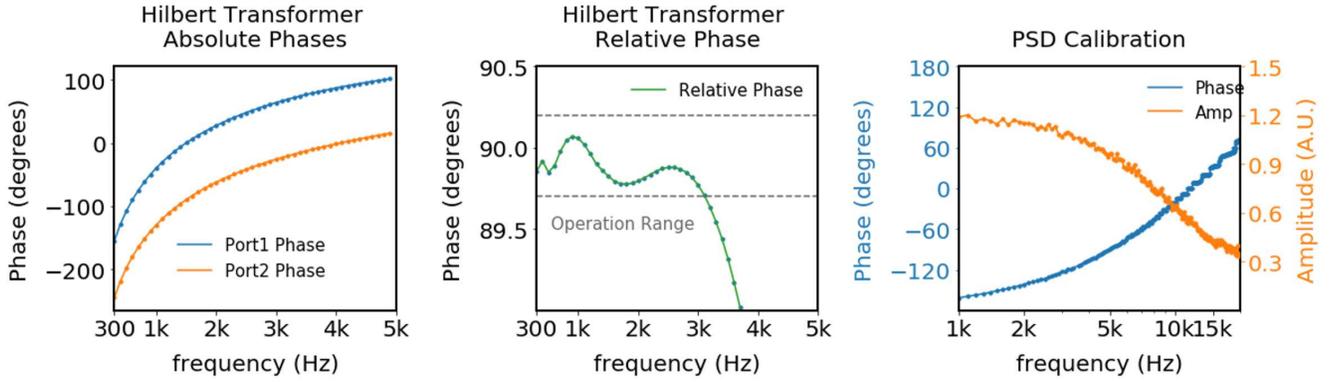}
\end{center}
   \caption{(a) Phase response of port1 and port2 on Hilbert transformer boards. (b) Relative phase between the two ports are within $90\pm0.3$ degrees in the operation range of $300-3000$ Hz. This range can be adjusted through changing resistance values in the circuit. (c) Phase and amplitude response of position sensor (PSD).}
\label{appendix_calib}
\end{figure}

%\label{si:detail_exp}
Phase calibration for elements in the control system is required for compensation during the system operation. Calibrations on Hilbert transformer board and position sensor (PSD) are shown in Figure~\ref{appendix_calib}. All calibrations are conducted with a high speed oscilloscope. As shown in Figure~\ref{appendix_calib}(b), the relative phase between port1 and port2 of Hilbert transformer board is within $90 \pm 0.3$ degrees range in the operational range of $300-3000$Hz.

\subsection*{Phase uncertainty of MEMS scanner}
\begin{figure}[t!]
\begin{center}
% \fbox{\rule{0pt}{3in} \rule{0.9\linewidth}{0pt}}
\includegraphics[width=1.0\linewidth]{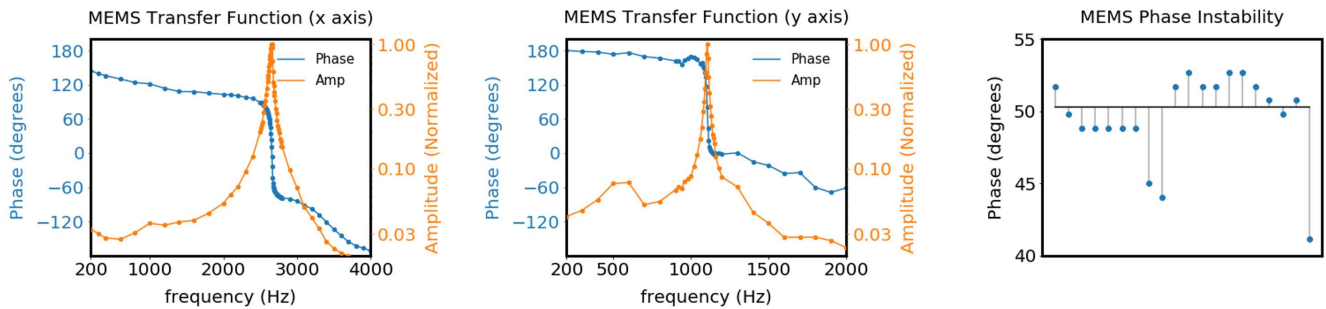}
\end{center}
   \caption{(a),(b) Transfer function amplitudes and phases for the used MEMS scanner. Quality factors $Q_x \sim 30$ and $Q_y \sim 50$ can be estimated from full-width-half-maximum (FWHM) on the $H_x$, $H_y$ curves. (c) Phase uncertainty of MEMS scanner. Relative phase between MEMS scanner motion and input actuation signal can be unstable due to random fluctuations in the system. An example is shown with $2660$ Hz actuation. Relative phase changes can be as large as $\sim 10^{\circ}$ within 40 minutes time range.}
\label{appendix_stable}
\end{figure}

%\label{si:instability}
Amplitudes and phases of transfer function $H_x(f_x)$, $H_y(f_y)$ are shown in Figure~\ref{appendix_stable}(a),(b), for two scanning axis of the MEMS scanner. From the calibration, resonant frequency of MEMS scanner is determined to be $f_x^r = 2660Hz$, $f_y^r = 1100Hz$. Quality factors are determined from full-width-half-maximum (FWHM) on the transfer function curve, $Q_x \sim 30$, $Q_y \sim 50$. As shown in Figure~\ref{appendix_stable}(a),(b), around resonance, relative phase between MEMS scanner motion and the input actuation signal undergoes a steep transition. Small fluctuations in the resonant frequency would result in large phase changes. This small fluctuation can be due to temperature fluctuations, spring stiffening and other random environmental factors.

We characterize this phase uncertainty of MEMS scanner when control system is NOT used. We record the relative phase between the x-axis actuation signal and scanner motion within 40minutes. The MEMS scanner is actuated at a fixed frequency $2660$ Hz. No modulation or control are used. Since the phase of actuation signal (from signal generator) is assumed to be stable enough, we attribute the $\sim 10^{\circ}$ relative phase change, shown in Figure~\ref{appendix_stable}(c), to fluctuations in MEMS scanner.

\subsection*{Power consumption}
The power consumption of proposed MEMS scanner system can be divided into two parts: control circuit power consumption and MEMS actuation power consumption. Here we give a detailed calculation:
\newline
The proposed control circuit consists of 4 operational amplifiers (Op-Amps), 2 for each scanning direction control. This contributes the major power consumption for the control circuit. During operation, each Op-Amp consumes $\sim V_{cc} \times I_{cc}$ power, where $V_{cc}$ is the supply voltage and $I_{cc}$ is the drain current. To achieve a low power consumption, Op-Amp such as TL062 can be used, with drain current as low as 200$\mu$A while still operates up to +30 Volts single supply.

The power consumption of a MEMS device can be expressed as $2 \pi fCV_{rms}^2$, where $f$ is actuation frequency, $C$ is the capacitor of MEMS and $V_{rms}$ is the RMS(root-mean-square) actuation voltage. Since an electrostatically actuated MEMS device usually has a capacitor $< 1$pF, and the actuation voltage is in the order of 10-100V, the MEMS device consumes a very small amount of power, in the order of nW to $\mu$W. Therefore, to apply such a scanner on portable/mobile devices, the major physical constraint is the maximum amplitude of actuation voltage, instead of the power consumption. When operating off-resonance and maximum actuation voltage is fixed, the scanning range decreases. This physical constraint is ignored in previous designs and motivates a more complete one proposed in this paper.

\end{document}